\documentstyle[11pt,aaspp4,epsfig]{article}

\newcommand{\Lam}{\Lambda}

\newcommand{\hinv}{{\rm h^{-1}}}

\newcommand{\kpc}{{\rm\,kpc}}
\newcommand{\himpc}{h^{-1}{\rm\,Mpc}}
\newcommand{\hikpc}{h^{-1}{\rm\,kpc}}
\newcommand{\kmsmpc}{{\rm\ km\ s^{-1}\ Mpc^{-1}}}

\newcommand{\Gyrs}{{\rm Gyrs}}
\newcommand{\yrs}{{\rm yrs}}
\newcommand{\yr}{{\rm yr}}

\newcommand{\T}{{\rm T}}
\newcommand{\K}{{\rm K}}

\newcommand{\Msun}{M_{\odot}}
\newcommand{\ang}{{\rm \AA}}

\newcommand{\vv}[1]{{\bf #1}}

\newcommand{\etal}{et~al.}

\def\eg{{\frenchspacing e.g.}}

\def\ltsim{\lesssim}
\def\gtsim{\gtrsim}

\def\bgeq#1{\begin{equation}\label{#1}}
\def\endeq{\end{equation}}
\def\bgeqa#1{\begin{eqnarray}\label{#1}}
\def\endeqa{\end{eqnarray}}

\def\Fig#1{Figure~\ref{#1}}

\begin{document}

\title{Luminosity Density of Galaxies and Cosmic Star Formation Rate 
from $\Lam$CDM Hydrodynamical Simulations}

\author{Kentaro Nagamine}
\affil{Joseph Henry Laboratories, Physics Department, Princeton University, 
Princeton, NJ 08544}
\author{Renyue Cen and Jeremiah P. Ostriker}
\affil{Princeton University Observatory, Princeton, NJ 08544}
\affil{(nagamine, cen, jpo)@astro.princeton.edu}

\begin{abstract}
We compute the cosmic star formation rate (SFR) and the rest-frame 
comoving luminosity density in various pass-bands as a function of 
redshift using large-scale $\Lam$CDM hydrodynamical simulations
with the aim of understanding their behavior as a function of redshift.
To calculate the luminosity density of galaxies,
we use an updated isochrone synthesis model which takes 
metallicity variations into account.
The computed SFR and the UV-luminosity density have a steep rise from 
$z=0$ to 1, a moderate plateau between $z=1 - 3$, and a gradual  
decrease beyond $z=3$. 
The raw calculated results are significantly above the observed 
luminosity density, which can be explained either by dust extinction or
the possibly inappropriate input parameters of the simulation. 
We model the dust extinction by introducing a parameter $f$;
the fraction of the total stellar {\it luminosity} (not galaxy population) 
that is heavily obscured and thus only appears in the far-infrared (FIR) 
to sub-millimeter (submm) wavelength range.
When we correct our input parameters (baryon mass-density and 
the yield of metals) to the current best-estimate, and apply dust extinction 
with $f=0.65$, 
the resulting luminosity density fits various observations reasonably well,
including the present stellar mass density, the local B-band galaxy luminosity
density, and the FIR-to-submm extragalactic background.
Our result is consistent with the picture that $\sim2/3$ of the total stellar 
emission is heavily obscured by dust and observed only in the FIR.
The rest of the emission is only moderately obscured which can be observed
in the optical to near-IR wavelength range.
We also argue that the steep falloff of the SFR from $z=1$ to 0 is partly due 
to the shock-heating of the universe at late times, which produces gas which 
is too hot to easily condense into star-forming regions.
\end{abstract}


\newpage

\section{Introduction}
\label{introduction}
We normally detect galaxies by observing luminous stars. Although luminous 
stars are only one of the constituents which make up galaxies, 
we hope to understand their internal properties and their evolution 
from the observations of luminous stars. The ultimate goal is to 
map out the star formation history of the universe from the 
end of the cosmic `dark age' to the present epoch, and constrain 
the growth of cosmic structure.

One then starts off by observing galaxies first in the 
optical wavelength where the atmospheric windows are open 
to ground-based telescopes, and move on to other 
wavelengths aiming to map out the entire electromagnetic spectrum.
An obvious way to estimate the SFR
from optical observations is to take advantage of the 
fact that the bulk of the rest-frame UV-emission from galaxies is 
dominated by the short-lived high-mass stars, and to assume
that the cosmic SFR is proportional to the rest-frame UV-luminosity 
density of galaxies at each epoch of the universe.

Many measurements of various SFR indicator at different redshifts 
have been published within the past few years, 
and our understanding of the cosmic SFR is evolving rapidly.
\cite{Lilly96}~(1996, hereafter L96) have shown that the SFR 
appears to be rising 
rapidly from $z=0$ to 1 as a function of redshift 
using the Canada France Redshift Survey (CFRS).
\cite{Madau96}~(1996, hereafter M96) used the Hubble Deep Field 
(HDF, \cite{Williams96}~1996)
to infer the SFR at high-redshift and obtained the SFR at $z=3$. 
They further went on to obtain the SFR at $z=4$,
which suggested that the SFR could be decreasing towards high-redshift 
($z\gtsim 3$; \cite{Madau97}~1997, hereafter M97).
This result, combined with the studies at $z=0 - 1.5$ 
(\cite{Gallego96}~1996; \cite{Connolly97}~1997, hereafter C97; 
\cite{Treyer98}~1998; \cite{Tresse98}~1998),
suggested that the SFR may have a peak at $z=1\sim2$, 
and spurred the excitement that we might have detected the peak 
of the star formation activity
in the history of the universe. The figure plotting the SFR as a 
function of redshift has since been called as the 
``Madau plot'' and gained popularity.
On the other hand, \cite{Sawicki97}~(1997) and \cite{Pascarelle98}~(1998)
obtained higher SFRs than M96 and M97, and suggested that the SFR
might be constant beyond $z\simeq 1.5$.
The most recent publication of the SFR at high-redshift ($z\gtsim 3$)
comes from \cite{Steidel99}~(1999, hereafter S99) who used 
the Lyman-break technique to detect galaxies at $z \sim 3$ to 4. 
They found more star formation at $z \gtsim 3$ than M96 and M97 did. 
They have also shown that the work of M96 and M97
might have missed some fraction of the galaxies due to an 
inappropriate color-cut on the color-color plane of the HDF galaxies. 
Thus, the results of M96 and M97 should be regarded as  
lower limits to the true SFR.

However, the discussion does not end here. In addition to 
the observational difficulty in determining the faint end of the 
luminosity function which makes the resulting SFR unreliable, 
the UV-emission from high-mass stars is
significantly obscured by dust which exists in and around 
the galaxies. Dust absorbs the UV-emission from high-mass stars
and re-emits the energy at longer wavelengths, FIR to submm. 
Thus the SFR inferred from the UV-emission detected in the
optical observations is only a part of the true star formation,
i.e., there is significant ``hidden'' star formation.

To reveal this hidden portion of the true star formation, 
observers have been making great progress using the 
Sub-millimeter Common User Bolometer Array (SCUBA) on 
the James Clark Maxwell Telescope (JCMT) at Hawaii.
The SCUBA observations at $850\mu m$ have already revealed 
hidden star formation
which was previously not detected by the UV-observations, and suggest
higher SFR in the past where star formation was more active than
the present. However, a quantitative deduction of hidden SFR 
is still difficult from their observations due to the limited 
spatial resolution of the telescope 
(\cite{Blain99}~1999; \cite{Trentham99}~1999).
Further observations in the infrared wavelength are
also important. We will discuss this issue further in \S~\ref{dust}.

In order to gain a better understanding of the physical mechanism
that affects the SFR, several groups have approached this subject 
from the theoretical side using dissipationless N-body simulations 
and semi-analytic models of galaxy formation
(\eg~\cite{White91}~1991; \cite{Kauffmann93}~1993; \cite{Baugh98}~1998, 
hereafter B98; \cite{SP98}~1998, hereafter SP98).
They have succeeded in drawing a picture 
roughly consistent with observations by using an appropriate 
stellar evolution model 
 and exploring the luminosity density as a function of redshift.
However, in these semi-analytic models, one has to follow the 
dissipative physical processes by putting in a model under 
simplified assumptions for the dynamics of galaxy formation. 
As \cite{Kauffmann98}~(1999) and SP98 note, there are significant 
uncertainties in their modeling of star formation and supernova 
feedback in the sense that the final observable outcome changes 
significantly by changing the free parameters in the model
(see SP98 for the list of free parameters and a nice comparison 
of different models). 
This is of course due to our ignorance of these complicated 
dissipative processes that take place during galaxy formation. 
One additional advantage of the semi-analytic 
approach is that it can search through a vast parameter space in 
a moderate amount of time.

 Although the basic uncertainty of the fundamental parameters 
in modeling these processes is similar, 
it is useful to carry out a detailed hydrodynamical simulation 
because it treats the dynamics of the products of these dissipative 
physical processes directly by following the behavior of the 
baryonic gas dynamically. 
Many important physical processes which we might not understand well enough to 
include in heuristic treatments (\eg~relative motion of 
dark matter and baryonic components) are included automatically when
the full set of equations is solved.
In particular, in semi-analytic models, the intergalactic medium or 
outskirts of galaxies are treated inaccurately in the sense that 
the shock-heating in these regions is largely ignored.
The simulations we use here are specially designed to capture the shock
regions well using the Total Variation Diminishing method (\cite{Ryu93}~1993).
However, due to its high computational requirement, 
the resolution of our large-scale hydrodynamical simulations has not 
reached the same level of that of the dissipationless dark matter 
simulations yet. Thus, the above two approaches are both important and 
complementary to each other. 


In this paper, we present the results of the computation of the cosmic SFR and 
the comoving rest-frame luminosity density of galaxies at various wavelengths
(1500$\ang$, 2800$\ang$, U, B, V, I, K-bands)
using a $\Lam$CDM hydrodynamical simulation
(\cite{accuracy}~1999a,b,c,d, hereafter CO99a,b,c,d).
Details of the simulation are explained in \S~\ref{simulation} and in 
Appendix~\ref{criteria}, indicating improvements over prior versions.

 In \S~\ref{sfr}, we calculate the cosmic SFR by simply 
summing up the mass of the collapsed baryonic objects formed 
in the simulation within a given time according to the star 
formation recipe explained in \S~\ref{simulation} and 
in Appendix~\ref{criteria}. 
We note that the SFR is a straight-forward and direct outcome of 
the star formation recipe and the input parameters that we implement 
in the simulation.

 In \S~\ref{lumdensity}, we calculate the comoving rest-frame 
luminosity density using 
the updated isochrone synthesis code by \cite{BC99}~(1993,1999, 
hereafter BC99) which takes the metallicity variation into account. 
Note that many of the earlier works which used population 
synthesis models have often assumed a constant solar 
metallicity in their calculations.

 In \S~\ref{discussion}, we consider possible corrections that 
can be applied to the raw result of our calculation to compensate for 
the inappropriate input parameters and the effect of dust extinction.
The first correction (\S~\ref{baryon}) 
is for the baryonic mass-density $\Omega_b$ which will be 
matched to the current best estimate value.
The second (\S~\ref{yield}) is the correction of the yield of metals. 
We adjusted the value so that the simulations reproduce the observed
intra-cluster gas metallicity.
The third (\S~\ref{dust}) is the dust extinction correction
to account for the hidden SFR. 
For the purpose of illustration, we consider a very simple 
model of dust extinction by introducing a parameter $f$ which is 
the fraction of the total {\it luminosity} 
that is heavily obscured: the entire optical emission from stars of 
this fraction is completely absorbed by dust and re-radiated 
in the FIR-to-submm wavelengths.
Note that this is {\it not} the fraction of the {\it galaxy population}, 
we are only referring to the {\it luminosity} itself 
for reasons to be explained in \S~\ref{dust}.
As the aim of this paper is not to construct a detailed model 
of dust, we satisfy ourselves by seeing that the calculated 
intrinsic luminosity density can be corrected to the observed
one with a reasonable value of this parameter $f$.
We refer to various observations to obtain a reasonable value of 
$f$ such as the local B-band luminosity density and the 
extragalactic background light.

Finally,  we consider the physical origin of the computed 
``Madau plot'' in \S~\ref{madauplot} and discuss the 
possible effect of the resolution in \S~\ref{resolution},
then conclude in \S~\ref{conclusion}.

\section{The Simulation}
\label{simulation}
The hydrodynamical cosmological simulations that we use here are 
also described in detail in CO99a,b,d; they are similar to but 
greatly improved over those of \cite{CO92}~(1992a,b). 
Motivated by \cite{Ostriker95}~(1995), the adopted cosmological 
parameters are $\Omega_m=0.37$, $\Omega_{\Lambda}=0.63$, 
$\Omega_b=0.049$, $\sigma_8=0.8$, $h=0.7$ where $H_0=100h\kmsmpc$,
and 25$\%$ tensor mode contribution to the cosmic microwave background
fluctuations on large scales. 
This model is consistent with the recent high-redshift
supernova observations which suggest the existence of the 
cosmological constant $\Omega_{\Lam} \simeq 0.7$
(\eg~\cite{Perlmutter98}~1998; \cite{Garnavich98}~1998).
The present age of the universe with these input parameters is 12.7 \Gyrs. 
We have two simulations with comoving box sizes of $L_{box}=(50,100)\himpc$,
each having $512^3$ cells and $256^3$ dark matter particles
with masses of $(6.6\times10^8, 5.3\times10^9)\hinv\Msun$, 
respectively. 
The {\it comoving} cell sizes are $(100,200)\hinv\kpc$, respectively.
At redshift 3, the mass resolution is unchanged but the spatial
resolution is $(25, 50)\hikpc$ in two boxes, respectively.
 
The code is implemented with a star formation recipe summarized in 
Appendix~\ref{criteria}.
It turns a part of the baryonic gas in a cell into a collisionless particle 
(hereafter ``galaxy particle'') once the following criteria are 
met simultaneously: 
1) the cell is overdense ($1+\delta_{tot} > 5.5$),
2) the cooling time of the gas $t_{cool}$ in the cell is shorter 
than its dynamical time $t_{dyn}$, 
3) the mass of the gas in the cell $m_{gas}$ is larger than 
the Jeans mass $m_J$, and
4) the gas flow is converging into the cell.

Each galaxy particle has a number of attributes at birth, including 
position, velocity, formation time, mass, and initial gas metallicity.
Upon its formation, the mass of a galaxy particle is determined by 
$m_{\ast}=0.25 m_{gas}\Delta t/t_{\ast}$ where $\Delta t$ is the current 
time-step in the simulation and $t_\ast={\mathrm max}(t_{\mathrm{dyn}}, 10^8\yrs)$.
The galaxy particle is placed at the center of the cell after its 
formation with a velocity equal to the mean velocity of the gas,
and followed by the particle-mesh code thereafter as collisionless 
particles in gravitational coupling with dark matter and gas.

The galaxy particles are baryonic galactic subunits with  
mass ranging from $10^3$ to $10^{10}\Msun$ (most of them having $10^{4-9}\Msun$), 
therefore, a collection of these particles is regarded as a galaxy. 
However, here we do not 
group them since we are only interested in the averaged
emissivity of those particles and grouping is irrelevant.
The number of galaxy particles at redshift zero is 
about $(19,25)$ million for the $(50,100)\himpc$ box.

Feedback processes such as ionizing UV, supernova (SN) energy, 
and metal ejection are also included self-consistently.
The SN and the UV feedback from young stars are treated as follows: 
$\Delta E_{SN} = m_{\ast} c^2 \epsilon_{SN}$
and $\Delta E_{UV}=m_{\ast}c^2 \epsilon_{UV} g_{\nu}$ where $g_{\nu}$ is
the Orion-like normalized spectrum of a young stellar association 
taken from \cite{Scalo86}~(1986), and ($\epsilon_{SN},\epsilon_{UV}$) are
the efficiency parameters taken as ($10^{-4.5}, 10^{-4.0}$), respectively 
(\cite{CO92}~1992a,b; \cite{CO93}~1993). 
The $\Delta E_{SN}$ is added locally in a cell, and 
the $\Delta E_{UV}$ is averaged over the box.
The metals are created according to $m_Z = Y m_{\ast}$ 
where $m_Z$ is the mass of the metals, $Y$ is the yield, 
and $m_{\ast}$ is the mass of the galaxy particle. They 
are ejected back into the inter-galactic medium and followed
as a separate variable by the same hydro-code which follows 
the gas density. The initial input parameter for the yield was 
$Y_{simu}=0.06$ which we will later correct to a lower value in  
\S~\ref{yield}.

\section{Cosmic Star Formation Rate}
\label{sfr}
\begin{figure}
\plotone{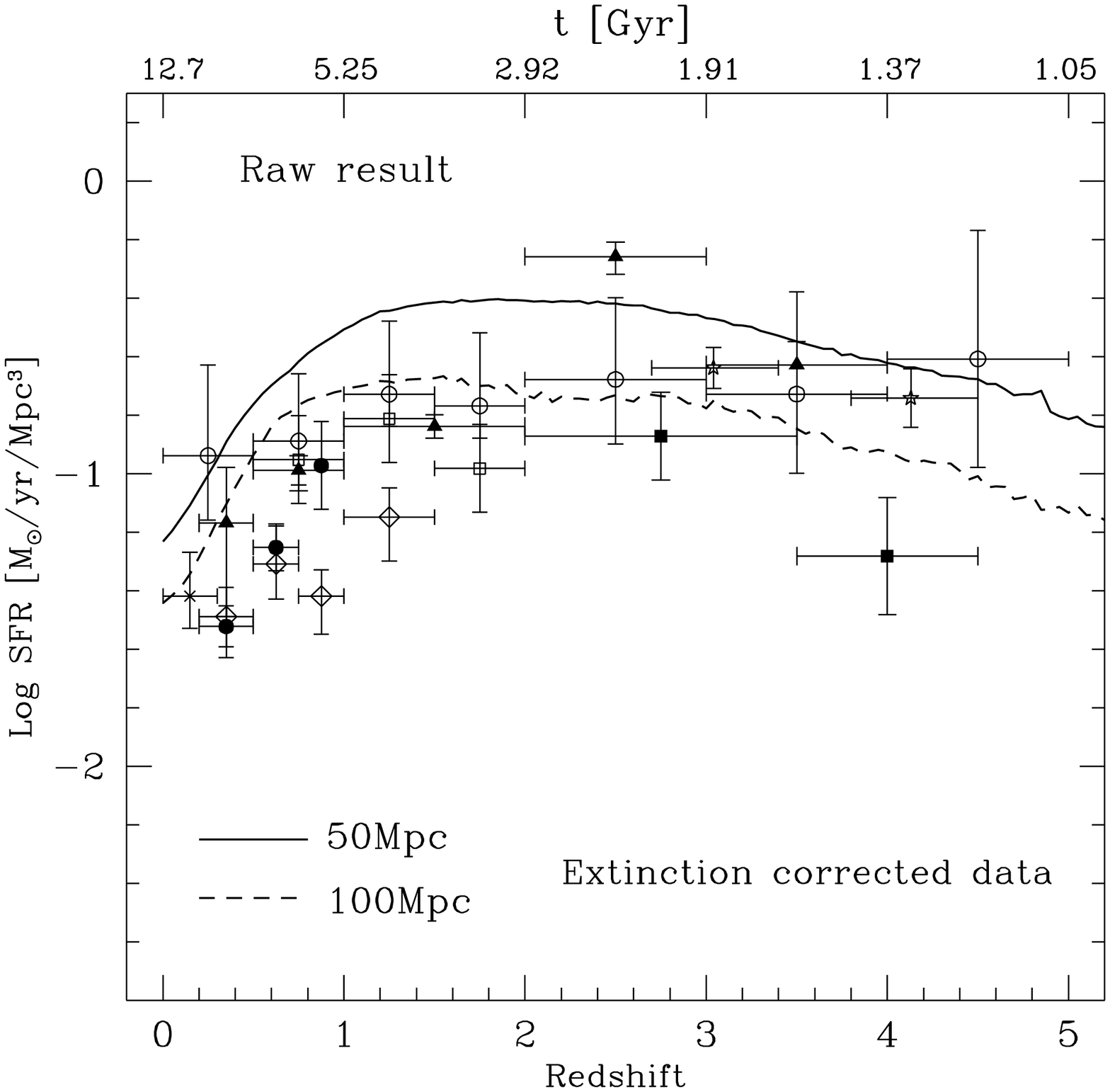}
\caption{{\it Raw} calculated SFR shown with the extinction corrected data points 
following the same prescription of S99 described in \S~\ref{sfr}.
The symbols for the data points are the same as shown in 
\Fig{fig2.ps}.
The data points are also corrected to our flat-$\Lam$ cosmology: 
$(\Omega_{m},\Omega_{\Lam})=(0.37,0.63)$ and $h=0.7$. 
The solid (50$\himpc$) and the short-dashed (100$\himpc$) 
lines are the results from the two simulation boxes.
\label{fig1.ps}}
\end{figure}

As we explained in the previous section and in Appendix~\ref{criteria},
stars are formed from baryonic gas as they cool and gravitationally collapse 
in the simulation. This recipe enables us to calculate the cosmic SFR 
as a direct output of the simulation; simply sum up the 
mass of the stars formed within a given amount of time.
Calculated in this manner, the SFR as a function of redshift is 
shown in \Fig{fig1.ps} (the `Madau plot').
Here we assumed that stars form instantaneously once a galaxy particle 
forms according to the recipe in Appendix~\ref{criteria}.

An alternative model is to assume that 
the collapsed baryons turn into stars gradually over the 
period of $\sim 10^{7-8}$ years in the spirit of \cite{Eggen62}~(1962). 
In this picture, the galaxy particle formed is still a mixture
of stars and dense gas clouds which quickly turns into stars
with the above time-scale.
However, the resulting SFR from this model is not very different 
from the instantaneous burst model except that the SFR at high redshift 
($z\gtsim 4$) is a little lower since some part of the star formation 
is delayed to later times.
Thus, we will assume the instantaneous starburst model hereafter;
that all of the mass of a galaxy particle turn into stars 
instantaneously upon its formation.

Together with the raw computed SFR from the two boxes, 
we also show the extinction corrected data points in 
\Fig{fig1.ps}. We corrected the original 
published values 
(L96, M96, C97, M97, \cite{Sawicki97}~1997, 
\cite{Pascarelle98}~1998, \cite{Treyer98}~1998, S99) 
by following the same prescription as S99, that is, to 
multiply by a factor of 4.7 (for $z\gtsim2$ data points) and 2.7 
(for $z\ltsim2$ data) according to the \cite{Calzetti97}~(1997) 
extinction law and use 
$\rho_{uv}[{\rm ergs/sec/Hz}]=8.0\times10^{27}$ SFR [$\Msun$/\yr] 
for the conversion from UV-luminosity density to SFR.
We note that \cite{Madau98}~(1998) obtained this 
conversion factor for the Salpeter 
initial mass function (IMF) with mass-cutoff of $[0.1, 125]\Msun$ 
at $1500\ang$ under the assumption of 
exponentially decaying SFR for each galaxy.
Here, the data points of S99 at $z\sim 3$ and 4 include 
an additional factor of 1.7 which should be included if one is to integrate 
the luminosity function all the way down to zero instead of 
down to $0.1L_{\ast}$ as explained in S99.
We have also corrected all the data points for our cosmology: 
from $(\Omega_{m},\Omega_{\Lam})=(1.0,0.0)$ (in which most observers 
present their results) to our $(0.37,0.63)$ case.
This correction shifts the data points downward overall,
but not more than 0.25 in ${\mathrm log}_{10}$, when the 
increase in both absolute luminosity and effective
comoving volume are taken into account for given apparent 
luminosity and number density of galaxies.
As a consequence, the rise in the luminosity density from $z=0$ to 1 
is suppressed compared to the Einstein-de-Sitter universe 
($\Omega_m=1.0$).

The two almost-parallel curves in \Fig{fig1.ps} 
are for the two different boxes ($L_{box}=50$ and 100$\himpc$) of 
the simulation; the curve of 50$\himpc$ box lies higher.
\cite{Weinberg99}~(1999) also observe that the 
amount of stars formed in similar simulations increases as they 
increase the resolution.
The results of the simulation should converge once we reach 
a sufficient resolution and better understanding of the 
physics of star formation. 
However, given the current level of sophistication 
of the simulation and our poor understanding of the star formation 
processes itself, it is clear that we are still far away from this
ideal goal. 
The total amount of stars created in the simulation also depends 
on several ingredients of the simulation, such as the feedback 
parameters described in \S~\ref{simulation} and the amplitude 
of the primordial power spectrum at small-scales.
Although the ambiguity remains, it is still worthwhile to 
discuss the mass-density of stars
$\Omega_{\ast}$ that were created in the simulation.
In our simulation, we find $\Omega_{\ast,raw}^{50}=0.018$ and 
$\Omega_{\ast,raw}^{100}=0.011$
where the upper index refers to the size of the simulation box
and the lower index ``raw'' represents that this is the raw result
without any corrections that may arise due to the inappropriate 
input parameters.
The observed $\Omega_{\ast}$ is still uncertain and ranges
from 0.002 to 0.006 in the literature (\eg~\cite{Fukugita99}~1999). 
Our raw result of the 
stellar mass-density may be too high, however,
this is the result under the assumption of $\Omega_b=0.049$.
We will discuss how this value would change if we were
to scale $\Omega_b$ to the current best estimate in 
\S~\ref{baryon}.
The steep falloff of the SFR from $z=1$ to 0 will be discussed in
\S~\ref{madauplot}.

\placefigure{fig1.ps}

\section{Comoving Rest-frame Luminosity Density}
\label{lumdensity}
We now calculate the emissivity of the galaxy particles 
using the instantaneous burst model of the isochrone 
synthesis code GISSEL99 (BC99). 
The input from the simulation is the formation 
time, the mass and the metallicity of each galaxy particle.
We calculate the emissivity of all the galaxy particles, 
sum them up at each epoch and divide by the total volume 
of the simulation box.
We do this exercise with three different IMFs; 
Salpeter (1955), Scalo (1986), and Miller-Scalo (1979).
We find that the Salpeter IMF produces the result that best matches 
the observations of galaxy luminosity density, so we will use 
the Salpeter IMF with mass-cutoff of $[0.1, 100]\Msun$ in the rest of this paper. 

\begin{figure}
\plotone{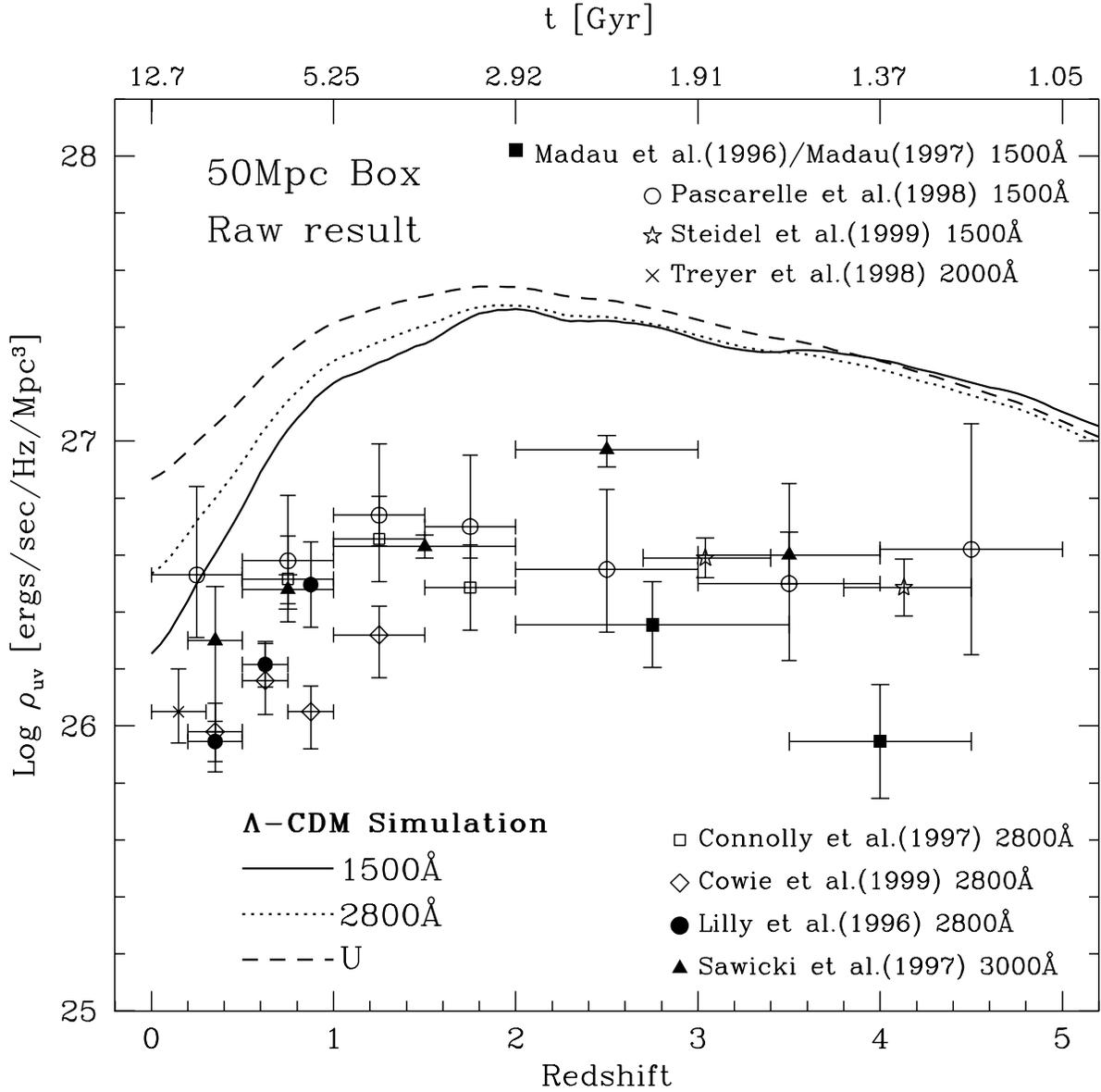}
\caption{{\it Raw} calculated comoving rest-frame UV-luminosity density of 
galaxies as a function of redshift. 
The data points are the observed galaxy luminosity density at the indicated
rest-frame wavelengths, corrected for our flat-$\Lam$ cosmology, 
but not for extinction.
\label{fig2.ps}}
\end{figure}

 In \Fig{fig2.ps}, we show the raw result of the 
calculated UV-luminosity density of the 50$\himpc$ box 
at 1500$\ang$, 2800$\ang$ and in the U-band. 
Naturally, the shape of the UV-luminosity density 
curve is similar to that of the SFR.
One should keep in mind that, as S99 point out, 
the error-bars of the high-redshift data points are 
under-estimated and will be much larger when the uncertainties
in the shape of the luminosity function at unobserved magnitudes 
is taken into account.

\begin{figure}
\plotone{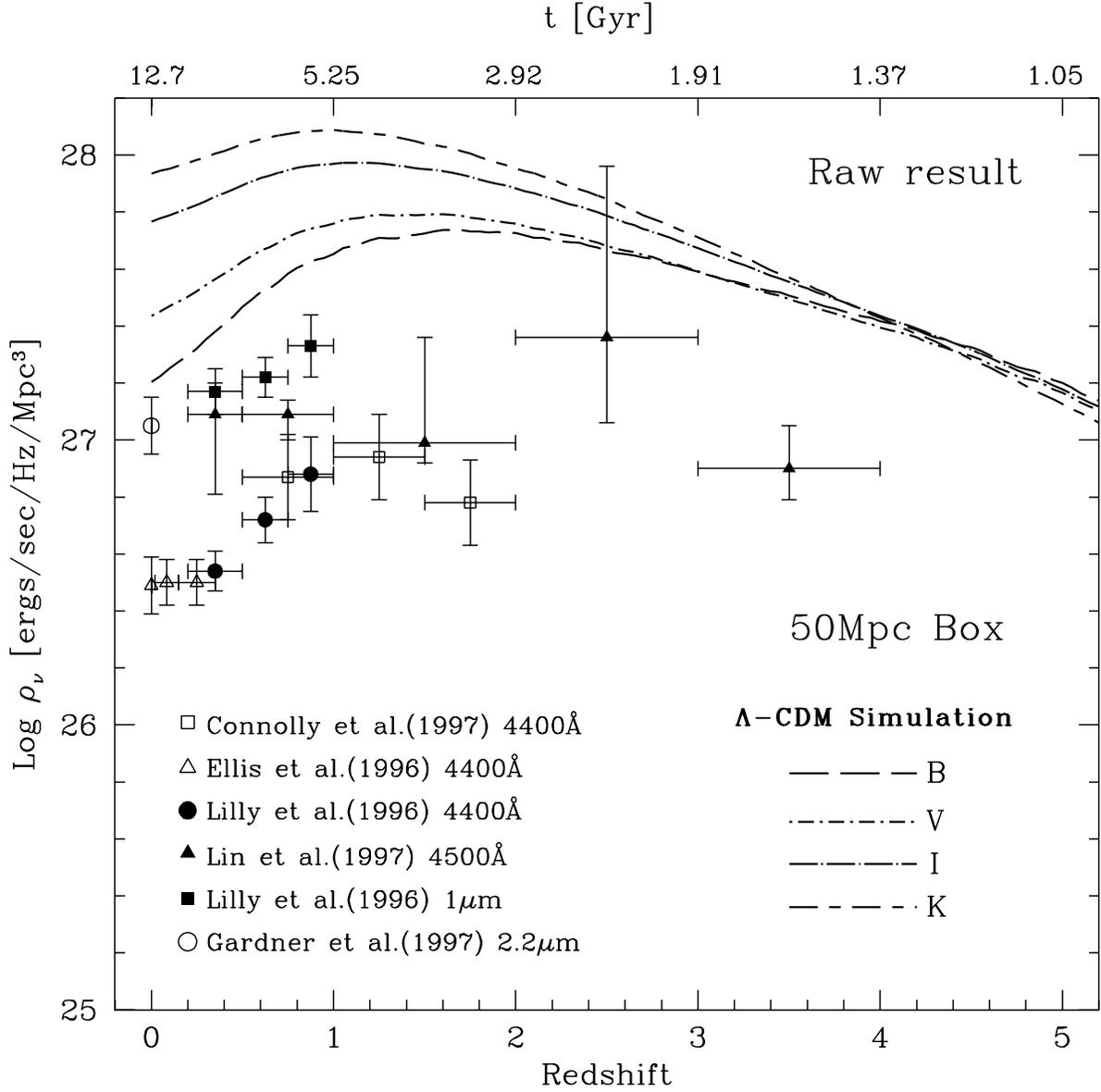}
\caption{Same as \Fig{fig2.ps}, but at longer wavelength. The data points are 
the observed galaxy luminosity density at the indicated rest-frame wavelengths, 
corrected for our flat-$\Lam$ cosmology, but not for extinction. 
\label{fig3.ps}}
\end{figure}

 Likewise, in \Fig{fig3.ps}, we show the raw result of 
the calculated galaxy luminosity density of the 50$\himpc$ box 
at longer wavelengths: B, V, I, and K-band. 
We plot the data points by \cite{Lin97}~(1997) who
re-analyzed the Hubble Deep Field (HDF) photometric redshift
data using 7 colors, whereas they only used 4 colors in their previous work
(\cite{Sawicki97}~1997). However, as they point out, the discrepancy 
between the results of the two analyses can be viewed as an indication of 
the systematic uncertainties involved in the analysis of this kind. 
All the data points in both figures are corrected for the adopted 
cosmology as we have explained in the previous section.

 Some of the previous works on the SFR limited the analysis
by assuming a constant solar metallicity for all galaxies
when calculating the luminosity density using population synthesis models.
We find that the inclusion of the metallicity gradient for 
galaxies is important to obtain the correct 
rate of decline in the luminosity density from $z=1$ to 0 in our simulation.
Therefore, the full dynamical treatment of gas and metals
in our simulation is an important advantage over the ad hoc 
treatment of these components in the semi-analytic models of 
galaxy formation.

In \Fig{fig2.ps} and \ref{fig3.ps}, 
the raw calculated result is significantly above the observations. 
We will now discuss the corrections 
that can be applied to the raw calculated result in the next section.

\placefigure{fig2.ps}

\placefigure{fig3.ps}

\section{Discussions}
\label{discussion}
In this section, we consider the corrections of our input parameters
of the simulation and the dust extinction which can be applied to 
the raw calculated results presented in the earlier sections. 
After explaining each correction, 
we show the corrected figures which are now consistent with
the observed luminosity density of galaxies in the UV-to-nearIR 
range.

\subsection{Correction of the Input $\Omega_b$}
\label{baryon}
The original input value of the baryon mass-density $\Omega_b$ for our
simulations was 0.049. The current best-estimate
of this value is 0.039 ($\Omega_b h^2=0.019$ with $h=0.7$; 
\cite{Burles99}~1999; \cite{Wang99}~2000).
So, how does the SFR change as a function of $\Omega_b$?

In our code, the radiative cooling rate is proportional to 
the square of the local baryonic mass-density times a function of 
temperature $T$ and metallicity $Z$: $dU/dt \propto \rho^2 f(T,Z)$ where
U is the internal energy of the gas. So, the cooled mass in our computation goes as 
$\Delta M_{\ast} \propto \rho^2 f(T,Z)\Delta V \Delta t$
where $\Delta M_{\ast}$ is the mass of stars produced, $\Delta$V is the 
volume element and $\Delta$t is the time step. That is, in a given time and 
volume element, the amount of mass turned into stars approximately scales as 
$\rho^2 \propto \Omega_{b}^2$. Thus, when we perform the simulation over again
in test cases using $\Omega_{b}=0.039$, rather than 0.049, we would have expected 
that the SFR would go down approximately by a factor 
$k_b \equiv (\Omega_{b,best}/\Omega_{b,used})^2=(0.039/0.049)^2=0.633$.  
This scaling is not exact because, in the higher $\Omega_b$ run, 
the metallicity is higher and cooling increases for that reason as well.
But this latter effect is small at the temperatures $\sim 10^4$ K and 
metallicities $Z/Z_{\odot}\ltsim 0.1$ which are most important ($Z_{\odot}=0.02$
is the solar metallicity). 

\begin{figure}
\plotone{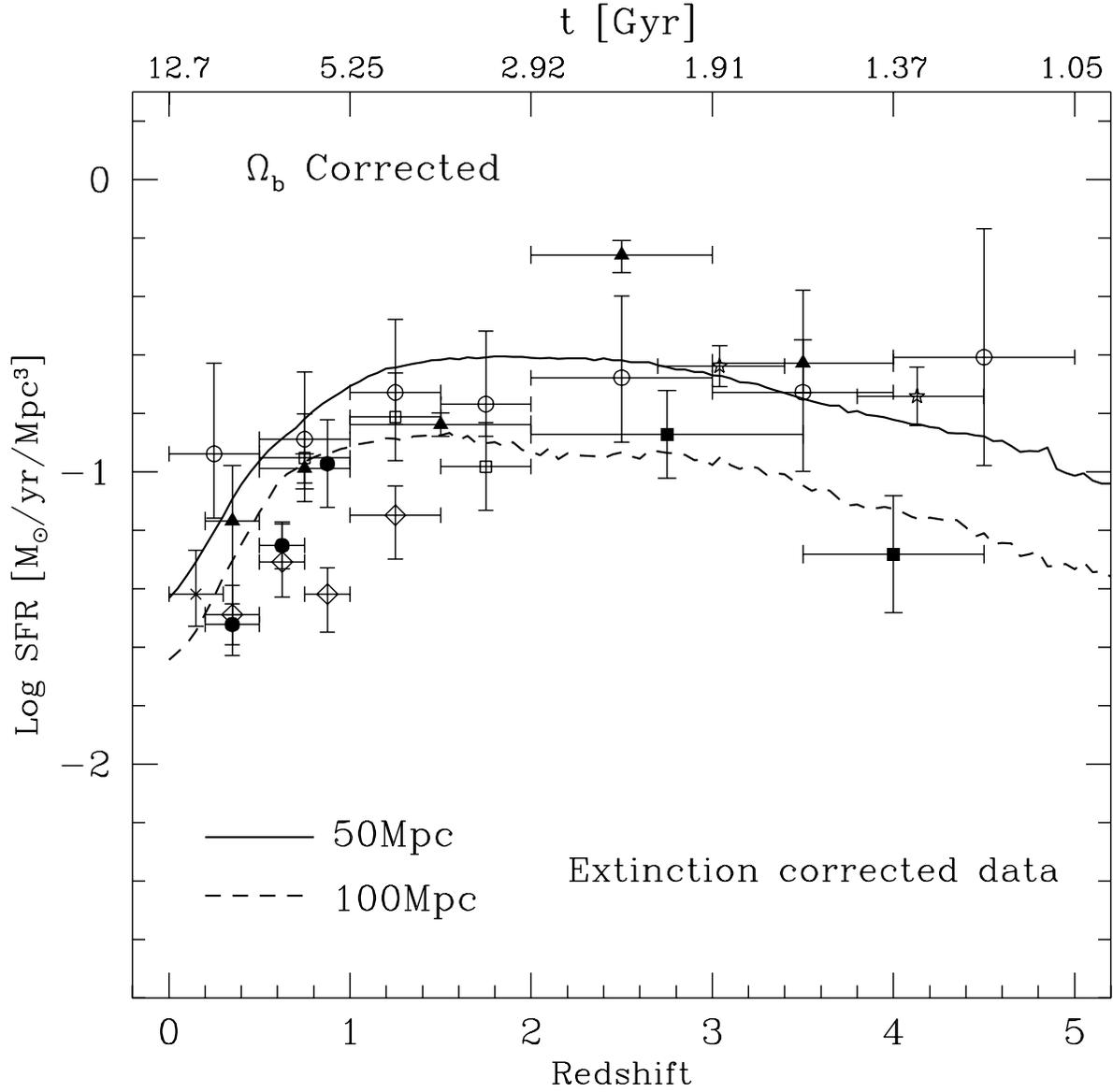}
\caption{Computed SFR after the $\Omega_b$ correction by a factor of $k_b=0.633$
compared to \Fig{fig1.ps} (see \S~\ref{baryon} for detail).
The same data points as in \Fig{fig1.ps} are shown.
\label{fig4.ps}}
\end{figure}

When the stellar mass is scaled with this factor $k_b$,
the calculated SFR shows a better fit with the observations in \Fig{fig4.ps}.
The calculated luminosity density by GISSEL99 also have to 
be scaled down by the same factor since it is linearly proportional to 
the stellar mass in the simulation (\Fig{fig5.ps} and \ref{fig6.ps} include
this correction). 
Accordingly, the stellar mass-density we discussed 
in \S~\ref{sfr} also goes down by the factor of $k_b$; 
$\Omega_{\ast,corrected}^{50}=0.012$ and 
$\Omega_{\ast,corrected}^{100}=0.007$ which are now closer
to the current best estimate ($\Omega_{\ast, obs}\ltsim 0.006$,
\cite{Fukugita99}~1999).
The remaining possible over-production of stars in our simulations can be 
explained by the high initial input value of the yield which 
will be discussed in the next subsection.

The $\Omega_b$ correction also affects the metallicity in the simulation 
since it changes the stellar mass by a factor of $k_b$ and 
the gas density by a factor of $\sqrt{k_b}$.
Thus, the metallicity, defined as the ratio of the amount of mass
in metals to the total baryonic mass for a given region, is
changed by a factor of $k_b/\sqrt{k_b}=\sqrt{k_b}$. This correction factor 
for the metallicity will be used in the next subsection.

\placefigure{fig4.ps}

\subsection{Correction of the Yield of Metals}
\label{yield}
Next, we consider the correction of the input value of 
the yield of metals. The original input value of the yield in our 
simulations was $Y_{simu}=0.06$ which simply was a typo-graphical error. 
We would like to scale this value so that we 
reproduce the observed value of the gas metallicity within clusters
of galaxies, namely $\frac{1}{3} Z_{\odot}$ (\cite{Mushotzky97}~1997). 
In order to identify clusters in the simulation, we grouped the galaxy particles 
using the HOP algorithm (\cite{Eisenstein99}~1999) and calculated the 
average gas metallicity within 1$\himpc$ around the center 
of the cluster. Taking only the clusters with 
galactic mass of $M_{galaxy}>10^{13}M_{\odot}$ where 
$M_{\odot}$ is the solar mass,
we found the mass-weighted mean of the intra-cluster gas metallicity 
to be $0.99Z_{\odot}$ for the 
50$\himpc$ box ($0.89Z_{\odot}$ for the 100$\himpc$ box). 
Thus, the correction factor $k_Z^{50}$ for the yield of 
50$\himpc$ box would be $k_Z^{50}=(0.33/0.99)=1/3$ which gives the 
the cluster normalized yield of 
$Y_{CLUSTER-NORM}^{50}=0.06 \times k_Z^{50} = 0.02$.

The yield correction of course changes the amount of metals itself.
Together with the $\Omega_b$ correction described in the previous 
subsection, the metallicity in the simulation has to be 
multiplied by a factor of $\sqrt{k_b} \times k_Z^{50}=0.27$. 
We have used the corrected metallicity in computing \Fig{fig5.ps} and 
\ref{fig6.ps}. These figures also include the effect of dust 
extinction which will be discussed in the next subsection.

There is one important caveat that goes with the change in the metallicity.
Since metals play a role in the cooling, it is related to the 
star formation criteria through the cooling time (see 
equation (A3) in Appendix~\ref{criteria}). More metals
make the cooling time shorter by efficient metal cooling,
thereby, resulting in the higher rate of star formation.
If the initial input value of the yield of metals was reasonably low, 
this was expected to be not a significant effect 
in the simulation because most of the galaxy particles 
form from the gas whose metallicity is lower than 
$\sim 0.1Z_{\odot}$ where metal cooling is not so important.
However, our initial input value of the yield was a little high
which may be the reason for the possible over-production 
of stars in our simulation compared to the current observational 
estimate of the stellar mass-density as we discussed in \S~\ref{baryon}.
This effect on cooling efficiency is complex, so we do not attempt to 
correct for this.

To be consistent in our computation of luminosity density,
we should use the same yield for both the simulation and GISSEL99.
Let us now estimate the yield assumed in the GISSEL99.
In the model of GISSEL99 we used, the Salpeter IMF with mass cutoff of
[0.1, 100]$M_{\odot}$ was assumed. By assuming the IMF, one is
setting the number of high-mass stars for a given total stellar mass, 
hence the yield as well. To make a rough estimate of the yield for this 
IMF, we refer to \cite{Arnett96}~(1996) where he calculates
the mass at least processed through helium burning.
Dividing this value by the integral of the Salpeter IMF in the 
mass-range specified above, we obtain the yield of this model 
of GISSEL99 to be $Y_{GISSEL}=0.016$. This value is close enough
to $Y_{CLUSTER-NORM}^{50}=0.02$ within the level of uncertainty
we are dealing with.

\subsection{Dust Extinction Correction}
\label{dust}
As we explained in \S~\ref{introduction}, a significant portion of 
the bolometric luminosity of galaxies is reprocessed by dust from
optical to FIR wavelength.
For the purpose of illustration, we consider a very simple model 
of dust extinction by introducing a parameter $f$ which characterizes 
the hidden fraction of the star formation. 
This is the fraction of the total bolometric luminosity of stars 
which is heavily obscured by dust (the effective optical depth in V-band
$\tau_{eff,V} \sim 100$) and only appears in the FIR. 
The rest of the fraction $1-f$ is only moderately obscured 
by dust ($\tau_{eff,V} \sim 0.2$ as suggested by \cite{Saunders90}~1990).

Note that we intentionally avoided calling $f$
the fraction of the galaxy population obscured by dust. 
It is often assumed that star formation 
takes place in two distinct populations of galaxies, one 
relatively unaffected by dust and the other completely obscured.
This is {\it not} the picture we have in mind here, we are only
referring to the fraction of the total {\it luminosity} emitted by stars.
This is because for example some galaxies may have a 
star-bursting region in its very dusty central core, 
and the rest of its body may contain only moderate amount
of dust. In fact, \cite{Adelberger00}~(2000) show that the 
majority of the star formation is taking place in galaxies
which contain moderate amount of dust, disfavoring the clear-cut
two distinct population model.
Our definition of the parameter $f$ strictly means the following: 
if we collect all the high-mass stars that are forming in the universe,
the fraction $f$ of them are in heavily obscured region and 
the rest of the fraction $1-f$ is in moderately obscured region
for which the obscuration is correctly estimated by Steidel and others.
Since the fraction of stars in heavily obscured regions contribute 
little in the UV and optical wavelengths, we must estimate their
numbers by other means.

As the aim of this paper is not to construct a detailed model 
of dust, we satisfy ourselves by seeing that the raw calculated
luminosity density can be corrected to the observed one with a 
reasonable value of $f$.
We now refer to various observations to infer the appropriate 
value of $f$, and discuss the consistency between them and our simulation.

\subsubsection{Low-redshift Starburst Galaxies}
\label{iso}
\cite{Calzetti00}~(2000) observed low-redshift starburst galaxies
using the Infrared Space Observatory (ISO) and found that 
only about 30$\%$ of the bolometric flux is coming out in the 
UV-to-nearIR wavelength range; the rest is emitted in the FIR.
This suggests $f\sim 0.7$ if their sample is typical in our universe.
They also found the following: 
1) the FIR spectral energy 
distributions of the starburst galaxies are best 
fitted by a combination of two modified Planck functions, 
with $T\sim 40-55K$ (warm dust) and $T\sim 20-23K$ (cool dust). 
The emission from this cold dust can be a major contributor 
to the FIR emission of starburst galaxies.
2) Because the wavelength coverage of the IRAS survey is 
$8-120\mu m$, the IRAS observations alone are not adequate 
for characterizing the emission from dust cooler than $T\sim30K$,
which could be a major contributor to the total dust mass content 
of a galaxy. On the other hand, the ISO satellite covers up to 
$\sim 240\mu m$ and thus is capable of revealing the FIR emission from 
the cold dust component. The resulting total FIR emission in the range of 
$1-1000 \mu$m is about twice larger than the IRAS FIR
emission in the range $40-120 \mu$m.

\subsubsection{Local Luminosity Density of Galaxies}
\cite{Saunders90}~(1990) estimated the FIR luminosity density of 
the IRAS galaxies to be $\langle L_{FIR, IRAS}\rangle = (5.6\pm0.6) \times 
10^7 h L_{\odot}Mpc^{-3}$. 
In order to compensate for the incomplete wavelength coverage of the 
IRAS survey, this value has to be scaled up by a factor of $\sim 2$
to obtain the correct FIR luminosity density according to 
\cite{Calzetti00}~(2000), which results in
$\langle L_{FIR, ISO}\rangle = (1.0 - 1.2)\times10^8 h L_{\odot}Mpc^{-3}$.

In our simulation, we find the bolometric luminosity density 
of galaxies to be $\langle L_{bol, simu}^{50}\rangle = 
5.8\times10^8 L_{\odot} Mpc^{-3}$
where the upper index denotes the size of the simulation box.
As an example, let us take $f=0.65$ and multiply this factor
to $\langle L_{bol, simu}^{50}\rangle$ to obtain the heavily 
obscured component which entirely goes into the FIR range:
we get $\langle L_{FIR, simu}^{50}\rangle = 3.8\times10^8 L_{\odot} Mpc^{-3}$.
Agreement within a factor of a few is adequate given the uncertainty 
in our model.

Another line of evidence for the high fraction of the hidden star 
formation relative to the visible one in the UV comes from the 
submm observations. Although quantitative constraints are still 
difficult to make from the submm observations as we described in 
\S~\ref{introduction}, a recent statistical analysis of the HDF 
observation in the submm range suggests that as much as 80$\%$ of 
the star formation could be hidden by dust (\cite{Peacock00}~1999).

For the rest of the fraction ($1-f=0.35$) which is only moderately obscured, 
we use a standard Galactic extinction curve of \cite{Draine84}~(1984) with 
standard MRN parameters, attenuating the light by a factor of $e^{-\tau_{eff}}$
where we take the effective optical depth at V-band $\tau_{eff,V}=0.2$. 
For this moderately obscured component, a consistency check 
with observation can be done with the local B-band galaxy 
luminosity density. \cite{Efstathiou88}~(1988) obtained 
$\langle L_{B,obs} \rangle =(1.9^{+0.8}_{-0.6})\times 
10^8 h L_{\odot}Mpc^{-3}$ 
using various galaxy surveys, and more recently \cite{Ellis96}~(1996)
got a consistent result from the Autofib Redshift Survey. 
In our simulation, we find
$\langle L_{B, simu}^{50}\rangle = 6.1\times 10^7 L_{\odot}Mpc^{-3}$ 
after applying the yield correction $k_{LUM}$ and the dust extinction
correction. Agreement within a factor of a few is adequate given the uncertainty
in our model.

\subsubsection{Extragalactic Background Light}
Another important observation that complements the observations
of the local universe is the extragalactic background light (EBL).
A photon of frequency $\nu^{\prime}$ emitted from a star at a certain 
redshift $z$ arrives at the earth with the redshifted frequency 
$\nu=\nu^{\prime}/(1+z)$. The entire history of the light emitted 
from stars in our universe is then recorded as the EBL:
\bgeq{EBLeq}
I_{\nu}=\frac{1}{4\pi}\int^{\infty}_{0}\rho_{\nu^{\prime}}\frac{dl}{dz} dz
\endeq
where $I_{\nu}$ [erg/sec/Hz/$\mathrm{cm^2}$/sr] is the EBL 
at redshifted frequency $\nu$, $\rho_{\nu\prime}$ is the rest-frame 
luminosity density at frequency $\nu$, and $dl/dz$ is the cosmological 
line element.
One can integrate $I_{\nu}$ with frequency and 
obtain the integrated EBL ($I_{EBL}$) 
often expressed in units of [nW/$\mathrm{m^2}$/sr] or 
as $\nu I_{\nu}$.

On the observational side, in the optical wavelength range, 
one can simply calculate the EBL from the observed 
fluxes of galaxies in the HDF. \cite{Pozzetti98}~(1998)
obtained the lower limit of the optical EBL using the HDF in the range 
of $3500-4500 \ang$. Later 
\cite{Madau00}~(2000, hereafter M00) extended the source of data
and obtained the optical EBL intensity of $I_{opt}\simeq 15$ 
nW/$\mathrm{m^2}$/sr. One should keep in mind that the estimates from 
the integrated fluxes from resolved galaxies will typically be
low because of the systematic errors (see M00), and must be strictly 
considered as lower limits.
 On the other hand, the direct observation by \cite{Bernstein99}~(1999) 
gives a total optical EBL intensity of $25 - 30$ nW/$\mathrm{m^2}$/sr
in the range of $3000-8000 \ang$ after the correction for the systematic
errors. This could even become $\sim 45$ nW/$\mathrm{m^2}$/sr
if the same systematic correction applies to the near-IR, and is 
higher than the integrated light from galaxies by a factor of $2-3$ (see M00).
Other constraints from the source-subtracted sky is discussed by 
\cite{Vogeley97}~(1997) and references therein.

In the FIR wavelength range, the COBE/FIRAS measurements in the 
$125-2000 \mu m$ range (\cite{Fixen98}~1998) and the DIRBE detection at 140 and
240 $\mu$m (\cite{Hauser98}~1998; \cite{Schlegel98}~1998)
suggest the approximate value for the FIR background is 
$I_{FIR}(140-2000\mu m) \simeq 20$ nW/$\mathrm{m^2}$/sr.
A study by \cite{Dwek98}~(1998) and DIRBE measurements at 2.2 and 3.5 $\mu$m
(\cite{Wright99}~1999) suggest $10-25$ nW/$\mathrm{m^2}$/sr in the range of 
$2.2-140\mu m$.
Combining all these measurements, M00 gave the `best-guess' 
total EBL intensity observed today as $I_{EBL}=55\pm20$ nW/$\mathrm{m^2}$/sr.

On the other hand, \cite{Finkbeiner00}~(2000) recently reported somewhat 
higher $I_{FIR}$ at 60 and 100 $\mu$m compared to the 
previous estimates from the analysis of DIRBE data-set, which may increase 
the contribution from the FIR range to the total budget of $I_{EBL}$. 
Another study by \cite{Gispert00}~(2000), using all the currently available
observational constraints, suggests the total EBL of $60-93$ nW/$\mathrm{m^2}$/sr. 
Both of these two recent works support the view that $\sim2/3$ of the total 
stellar emission is reprocessed by dust and emitted in the FIR range.

In our simulation, we obtain the total EBL of 
$I_{EBL, simu}=81$ nW/$\mathrm{m^2}$/sr after the corrections  
described in previous sections, consistent with the most recent estimate of 
\cite{Gispert00}~(2000).
Taking the value $f=0.65$ for the hidden SFR which entirely emits in 
the FIR-to-submm range, and with the moderate obscuration of 
$\tau_{eff,V}=0.2$ for the rest of the 35$\%$, 
our estimate from the simulation is $I_{opt,simu}\simeq 25$ 
nW/$\mathrm{m^2}$/sr and $I_{FIR,simu}\simeq 55$ nW/$\mathrm{m^2}$/sr. 
The numbers from the simulation and the observations are consistent with 
each other within their uncertainties.

\subsubsection{Summary on the Dust Extinction Model}
Assuming that 65$\%$ of the total stellar emission is heavily obscured 
and completely reprocessed by dust from optical to FIR wavelength, and 
that the rest of the 35$\%$ moderately obscured with the effective optical 
depth of $\tau_{eff,V}=0.2$,
we obtain the best match between our calculation and the observed 
UV-to-nearIR galaxy luminosity density.
Although the fit is not perfect, considering both the observational
and theoretical uncertainties, agreement within a factor of a few is adequate.

\begin{figure}
\plotone{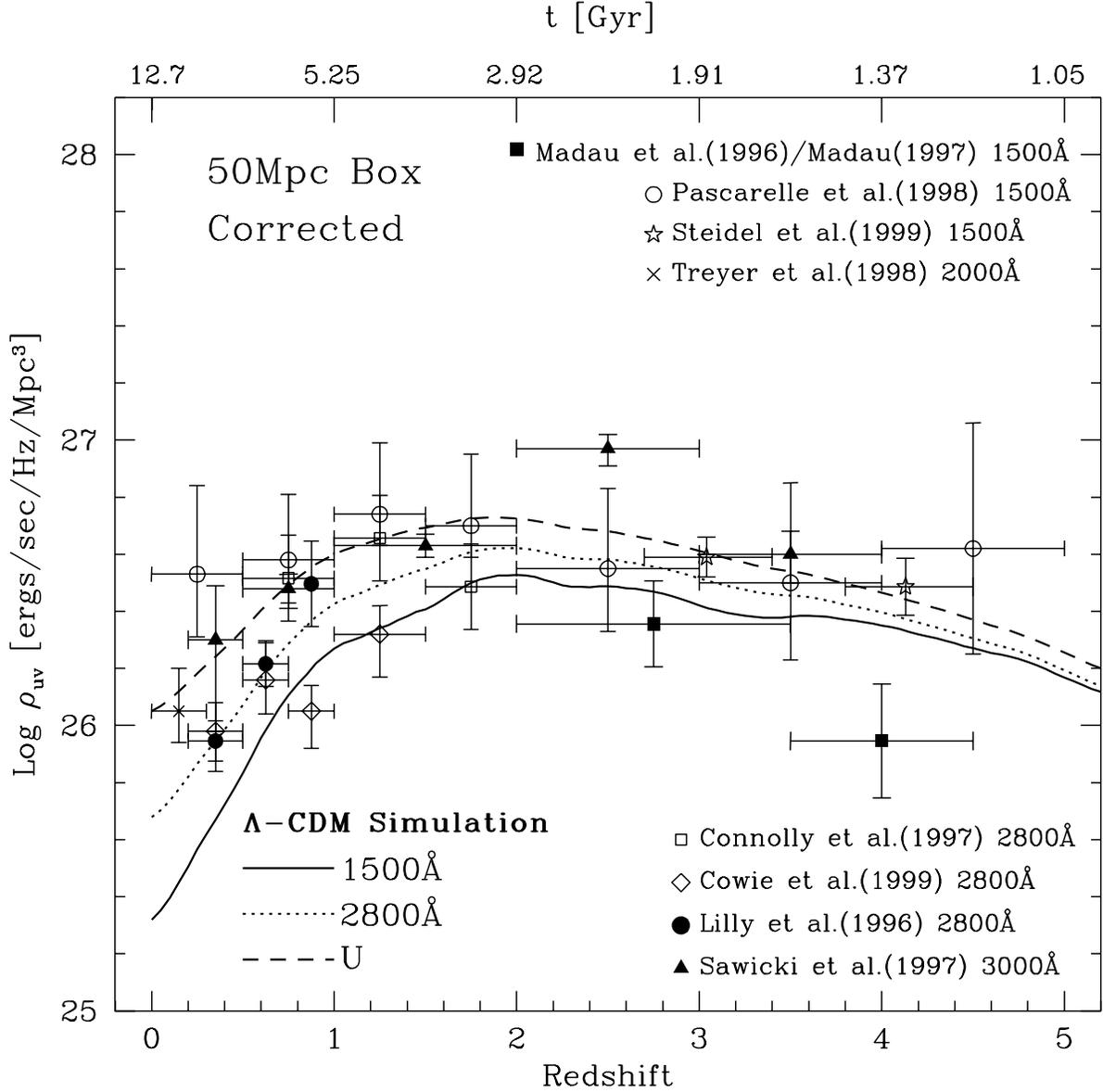}
\caption{Computed UV-luminosity density corrected for $\Omega_b$, yield and 
dust extinction (see \S~\ref{discussion} for detail).  
The same data points as in \Fig{fig2.ps} are shown.
\label{fig5.ps}}
\end{figure}

\begin{figure}
\plotone{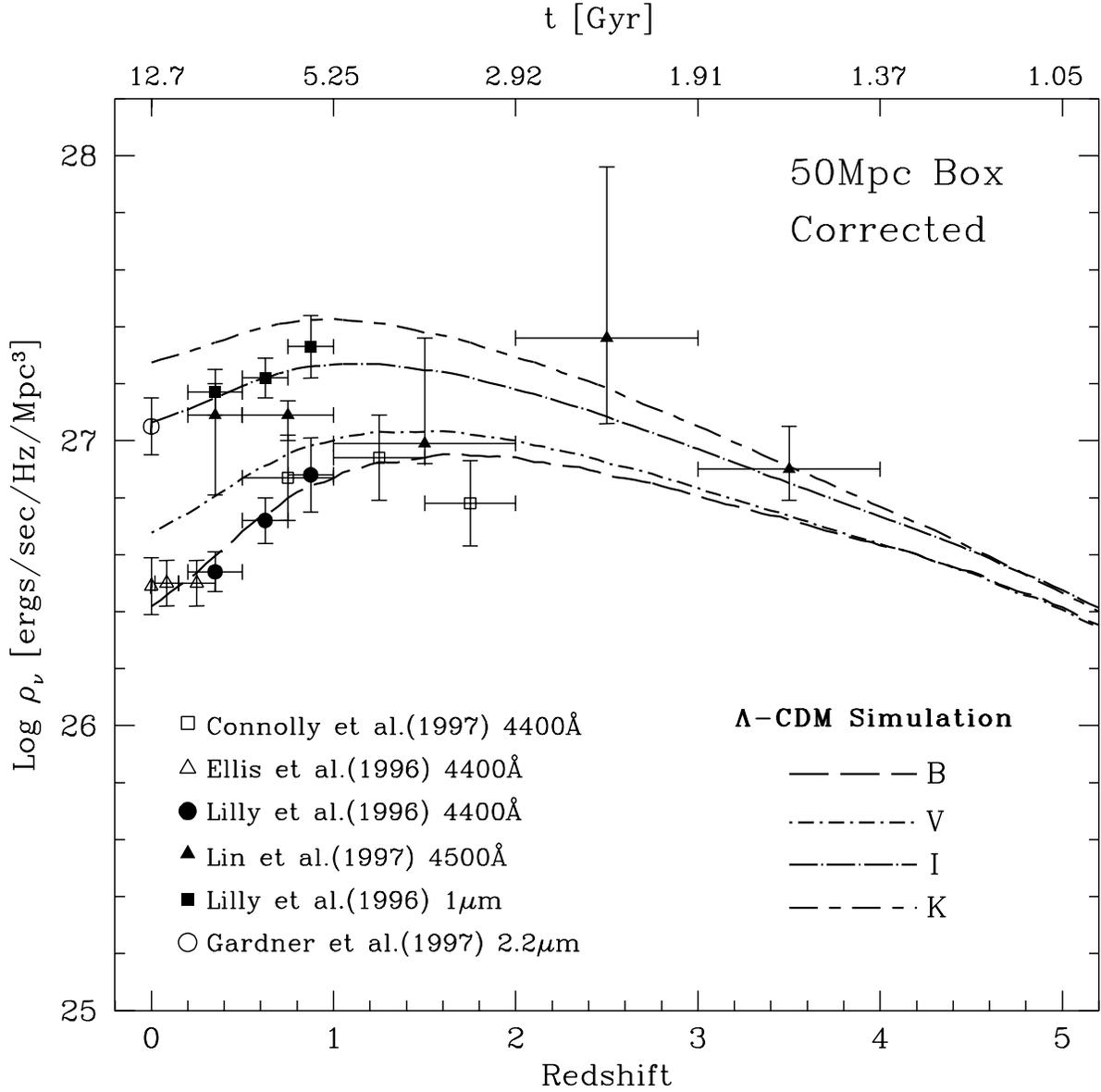}
\caption{Same as \Fig{fig5.ps}, but at longer wavelengths.
The same data points as in \Fig{fig3.ps} are shown.
\label{fig6.ps}}
\end{figure}

Using the corrected metallicity and the stellar mass as described in the 
previous subsections, we re-calculate the galaxy luminosity density from our
simulation and obtain \Fig{fig5.ps} and \ref{fig6.ps} after dust extinction
correction.
They are now in good agreement with the observational data shown.
Thus, our simulation is consistent with the picture that 
$\sim$ 2/3 of the total stellar emission is heavily obscured by dust,
and can only be observed in the FIR-to-submm.
The rest of the 35$\%$ is moderately obscured, and can be 
observed in the UV-to-near-IR range. 
A significant portion of the heavily obscured fraction of the total light
was previously missed by the IRAS, and it is now starting to be 
resolved by the ISO observations as explained in \S~\ref{iso}.

\placefigure{fig5.ps}

\placefigure{fig6.ps}

\subsection{On the Computed ``Madau Plot''}
\label{madauplot}
In this section, we discuss the falloff of the SFR from $z=1$ to 0.
This sharp dropoff of the SFR at low-redshift was first pointed out by L96. 
However, \cite{Cowie99}~(1999) argue that the L96 results has two 
weaknesses: first, it is a red selected 
sample which requires a substantial extrapolation to obtain UV-luminosity 
density; second, it does not sample enough of the luminosity function 
to allow reliable extrapolation to a total luminosity density.
They argue for a shallower decline of the SFR from $z=1$ to 0. 
Thus, the observational evidence of the sharp dropoff of the SFR
from $z=1$ to 0 is not yet conclusive, but the decline itself 
seems to be solid.

In \Fig{fig1.ps} and \ref{fig4.ps}, our simulations reproduce this sharp 
decline in SFR well. It has been known that in semi-analytic models it is 
difficult to reproduce this sharp dropoff from $z=1$ to 0 (\eg~B98; SP98).
We expect that this sharpness of the decline of SFR is at least 
partly due to the physical effect discovered by \cite{Blanton99}~(1999) 
in the same simulation. They found that at low-redshift the highest density 
regions cease to contain high star formation galaxies as they are 
typically in deep potential-wells where the high gas temperature 
prevents efficient star formation.

 Thus, the physical origin of the ``Madau feature'' (the rise
and fall of the SFR) can be explained as follows.
At high-redshift ($z\gtsim 1$), the gas temperature is still 
sufficiently low that the high-density regions are favored 
for star formation since they have shorter cooling time 
(note $t_{cool}\propto T/\rho \Lam$ where $\Lam$ is the cooling rate). 
As it approaches to low-redshift ($z\ltsim 1$), 
the gas temperature increases due to the shock-heating and the SN feedback 
in high-density regions. As a result, the cooling time in high-density 
regions gets longer and the star formation becomes inactive.
This naturally explains the rise and fall of the cosmic SFR, 
and is consistent with the result of our simulation.
We refer the readers to the paper by \cite{Blanton99}~(1999) 
for a more quantitative analysis.

 The findings of CO99a also support this idea.
Again using the same simulations, they found that the WARM ($\T<10^5\K$) 
gas which occupies 95$\%$ of the total baryonic mass at 
redshift 3 is heated up into HOT ($\T>10^7\K$; 20$\%$ at z=0) 
and WARM/HOT ($10^5<\T<10^7\K$; 50$\%$ at z=0) gas by 
shock-heating and SN feedback. 

 The above arguments lead to the idea of ``SFR as a function 
of overdensity''. We can demonstrate this idea in a straight-forward
way by making the Madau plot as a function of overdensity, which 
is presented elsewhere (\cite{Nagamine00}~2000).

 We note that our simulation adopts the flat-$\Lam$ cosmology
where there is more volume at moderate redshift and longer age 
of the universe. \cite{Totani97}~(1997) have pointed out that the 
$\Lam$-dominated flat model is favored to reproduce the 
steep dropoff of the UV-luminosity density below redshift 1.
The rate of increase of the SFR from $z=5$ to 2 in our simulation 
is similar to that of the semi-analytic treatments by B98 and SP98.

\subsection{Effect of Resolution}
\label{resolution}
In this paper, we focus on the quantities which are averaged 
over the simulation box so that our conclusions will suffer 
minimally from the resolution problem of simulations, 
and test the robustness of our conclusions
by comparing the results with those obtained using twice the spatial 
and eight times the mass resolution of the larger scale simulation.

One such test was performed in \cite{physicalbias}~(1999b) where 
they showed that the bias and the correlation function calculated 
from the two boxes does not differ more than 10$\%$ when smoothed 
with a Gaussian window of the same radius 0.5$\himpc$, 
on scales $1\himpc<\Delta r<10\himpc$. 
Therefore, they concluded that the finite numerical resolution 
does not significantly affect the results as long as the quantities 
averaged over a large scale are considered.

 However, the simulation probably under-estimates the star formation 
rate with limited resolution although it is not clear by how much.
On the other hand, we have over-estimated the star formation due to the
high initial input value of the yield which lead to too efficient metal cooling.
All these effects together, it is difficult to assess by how much 
we have over- or under-estimated the star formation overall.

 As for the redshift dependence of the resolution, 
our simulation has higher physical resolution at high-redshift 
relative to low-redshift as we described in \S~\ref{simulation}. 
Considering the fact that the higher resolution 
simulation produces more stars as described in \S~\ref{sfr}, 
we might be under-estimating the amount of stars at low-redshift
if the resolution affects the star formation seriously. 
Thus, we cannot completely rule out the possibility that 
the steep dropoff of the SFR from $z=1$ to 0 might have been 
enhanced due to the poorer resolution at low-redshift relative
to that at high-redshift. 
However, at the same time, we also see the physical effects 
found by \cite{Blanton99}~(1999) in the same simulation which 
encourages us to believe that we are capturing the essential 
features of what is happening in the real universe.
We hope to examine this issue more using the upcoming higher 
resolution simulations.

\section{Conclusions}
\label{conclusion}
We have presented the result of the computation of the cosmic 
SFR and the luminosity density of galaxies using large-scale 
$\Lam$CDM hydrodynamical simulations.

The raw calculated results are significantly above the 
observations (\Fig{fig1.ps}, \ref{fig2.ps}, and 
\ref{fig3.ps}).
However, we showed that the fit with the UV-to-nearIR 
observations of galaxy luminosity density
greatly improved when the corrections of the initial input parameters
and the dust extinction was applied (\Fig{fig4.ps}, 
\ref{fig5.ps}, and \ref{fig6.ps}).
The baryon mass-density was matched to the current best-estimate 
and the metallicity was corrected so that the observed intra-cluster gas 
metallicity was reproduced in the simulation. 
The resulting effects from these corrections
were discussed in detail by comparing with the currently available 
observations including the present stellar mass-density, the local B-band galaxy 
luminosity density and the extragalactic background light in the optical
and the FIR wavelengths.

Our dust extinction model includes a parameter $f$
which characterizes the fraction of the total stellar {\it luminosity} 
that is heavily obscured and thus only appears in the FIR-to-submm 
wavelengths (the ``hidden'' SFR). 
We found that the value of $f=0.65$ was preferred in order to 
obtain a good fit with the UV-to-nearIR observations of 
the luminosity density of galaxies.
The model suggests that as much as $\sim2/3$ of the total 
stellar emission is heavily obscured  ($\tau_{eff,V} \sim 100$) and 
the rest is moderately obscured ($\tau_{eff,V} \sim 0.2$).
A large portion of the heavily obscured fraction was previously 
missed by the IRAS, and it is now starting to be resolved by the recent 
ISO observations as explained in \S~\ref{iso}.
With the above parameters, we found that the computed optical and 
FIR extragalactic background light from our simulation are consistent 
with the current observational estimates within the observational
and theoretical uncertainties. 

The computed SFR after the $\Omega_b$ correction (\S~\ref{baryon}) 
agrees well with the extinction corrected data points (S99) of the optical 
surveys of galaxies (\Fig{fig4.ps}). Our result indicates the steep rise
of the SFR from $z=0$ to 1, a moderate plateau between $z=1$ to 3,
and a gradual decrease towards high-redshift ($z\gtsim 3$).
We argued in \S~\ref{madauplot} that the rapid falloff of the SFR 
from $z=1$ to 0 is partly due to the physical effect discovered 
by \cite{Blanton99}~(1999): the gas in high-density 
regions is heated by shock-heating and supernova feedback, 
and the star formation in those regions is suppressed at low-redshift
because the cooling time of the gas becomes longer due to the 
high temperature in the region.

\acknowledgements
We thank Stephane Charlot for providing us with the result of
the isochrone synthesis code GISSEL99, and Bruce Draine for providing us 
with the dust extinction curve.
We also thank Masataka Fukugita, Michael Strauss, Piero Madau, Michael Blanton,  
Joel Primack, and Charles Steidel for useful comments and discussions.
We are grateful to the anonymous referee whose report greatly 
improved the quality of this paper.
This work was supported in part by grants AST9318185 and ASC9740300.

%
%
\appendix
\section{Galaxy Particle Formation Criteria in the Simulation}
\label{criteria}

The criteria for galaxy particle formation in each cell of the 
simulation are:
\begin{eqnarray}
1+\delta_{tot} &>& 5.5, \\
m_{\mathrm{gas}} &>& m_{\mathrm{J}} \equiv G^{-3/2} \rho_{\mathrm{b}}^{-1/2}
C^3 \left[1 +\frac{1+\delta_d}{1+\delta_b} 
\frac{\bar{\rho_d}}{\bar{\rho_b}}\right]^{-3/2}, \\
t_{\mathrm{cool}} &<& t_{\mathrm{dyn}} \equiv \sqrt{\frac{3\pi}{32
G\rho_{tot}}} \mathrm{,~and} \\
\nabla\cdot\vv{v} &<& 0
\end{eqnarray}
where the subscripts ``$b$'', ``$d$'' and ``$tot$'' refers to baryon,
collisionless dark matter and the addition of the two respectively.  
$C$ in the definition of the Jeans mass is the isothermal sound speed.
The cooling time is defined as $t_{cool}={\rm k_B T}/{\rm \rho \Lam}$ 
where ${\rm \Lam}$ is the cooling rate in units of [${\rm erg~sec^{-1} cm^3}$]
and $\rho$ is the gas density. Other symbols have their usual meanings.
See \S~\ref{simulation} as well.

\clearpage

\clearpage

\clearpage

\clearpage

\clearpage

\clearpage

\end{document}